# Assessment of a hybrid software development process for student projects: a controlled experiment

Rafał Włodarski

Institute of Information Technology
Lodz University of Technology
Lodz, Poland

rafal.wlodarski@edu.p.lodz.pl

Jean-Remy Falleri

Univ. Bordeaux, Bordeaux INP, CNRS,
LaBRI, UMR5800, F-33400 Talence,
France. Institut Universitaire de France

falleri@labri.fr

Corinne Parvéry

Bordeaux INP
Talence, France

corinne.parvery@bordeaux-inp.fr

*Abstract*—In recent years, a vivid interest in hybrid development methods has been observed as practitioners combine various approaches to software creation to improve productivity, product quality, and adaptability of the process to react to change. Scientific papers on the subject proliferate, however evaluation of the effectiveness of hybrid methods in academic contexts has yet to follow. The work presented investigates if introducing a hybrid approach for student projects brings added value as compared to iterative and sequential development. A controlled experiment was carried out among Bachelor students of a French engineering school to assess the impacts of a given development method on the success of student computing undertakings. Its three dimensions were examined via a set of metrics: product quality, team productivity as well as human factors (teamwork quality & learning outcomes). Several patterns were observed, which can provide a starting point for educators and researchers wishing to tailor or design a software development process for academic needs.

## I. Introduction

As the software industry evolves to seize technological opportunities and respond to new challenges, so do the development approaches applied by practitioners. As a result, since the conception of the "Waterfall" model, a wide variety of software processes and life cycle models has been established, documented and applied in both commercial and academic contexts. Nevertheless, in many practical cases, a given methodology applied by the book does not address issues arising from a particular development environment.

A variety of publications, including scientific papers [1, 2] as well as more practitioner-oriented studies [3] show a trend towards the development and use of hybrid approaches. As defined by Kuhrmann et al. [4], these are "any combination of agile and traditional (plan-driven or rich) approaches that an organizational unit adopts and customizes to its own context needs (e.g. application domain, culture, process, project, organizational structure, techniques, technologies, etc.)". What can be observed is that plan-driven projects incorporate iterative development to accommodate change and introduce additional activities to generate more contact hours with users [5]. Likewise, agile-based methods integrate elements from "traditional" approaches such as architecture planning and formal estimation [5].

Research on the hybrid approaches is still relatively scarce, and that is particularly true when it comes to their application to drive student projects. To the authors' best knowledge, there are only two papers reporting on the use of a hybrid approach in the academic context [6, 7]. In this study, we aim to compare the effectiveness of a hybrid way of working against the processes it combines: iterative and sequential. We thus introduce simplified versions of iterative, sequential and hybrid methods to guide computing projects developed by teams of novice students and compare their outcomes.

A recent study [8] lists the goals most frequently named by practitioners devising hybrid methods; they can be regrouped into two broad categories: project quality (e.g. external product quality) and project efficiency (e.g. enhanced productivity and time-to-market). While both remain relevant for student undertakings, the ultimate goal of any course is fulfilment of its underlying learning outcomes. Therefore, in our study we evaluate three dimensions of success: project quality, team productivity and human factors (teamwork quality & learning outcomes). These elements are investigated to address the following research question:

**RQ**: Does a hybrid method yield better results in one of the evaluated success dimensions of student projects than the processes it combines (sequential and iterative)?

To answer the research question, a total of 16 metrics were evaluated for a group of 67 third year students of a Telecommunications program. We were also conducting post-experiment surveys in which we measure subjects' perception of the response variables in the human factors category. The remaining sections of the paper describe the experiment planning process and operational aspects of its execution. Results of the study are then presented, and their validity discussed. The paper concludes with a discussion on the findings and their contribution to the body of knowledge on software engineering education.

## II. Experiment planning

### A. The study environment – setting and artefacts

The context of the experiment is a Web programming course, a compulsory class worth 2.5 ECTS points (5 hours of lectures, 20 hours of tutorials, and 5 weeks x 2h of supervised assignment work). Following the European Credit Transfer and Accumulation System guidelines [9], students were expected to dedicate 30 to 40 hours on the course outside of activities guided by teaching staff.

The assignment was a PHP, HTML and CSS-based system to keep track of and share expenses with others, a concept similar to applications like Tricount and Splitwise. Due to a limited duration of the project, the requirements engineering activity was carried out prior to the course by its instructors and the students were provided with a Backlog of user stories, classified according to their priority:
- P1 – core of the application, to be implemented first
- P2 – major functionality, to be treated with high priority
- P3 – nice to have functions, to be developed if time allows

Given that the authors investigate a university context, typically characterized by stable requirements, an informed decision not to introduce changing needs was taken as it could skew the experiment towards the iterative approach.

As part of the assignment, all teams were required to provide the following artefacts:

1. application source code as well as its deployment script
2. a project documentation, containing the following sections:
- description, of the project context and its target users i.e. 'personas',
- sitemap of the application, indicating its GUI structure,
- wireframes of all the screens of the application to be developed,
- a list of the system modules, their description and underlying PHP scripts - a basis of the architecture
- a list of test cases as well as a requirements traceability matrix that links user stories to their tests, as well as tracks their execution and outcomes.

Detailed instructions and samples were provided to ensure proper understanding of what is expected and facilitate students' work.

### B. The study environment - processes

The course population was divided into three separate laboratory groups, where different processes were introduced: iterative, sequential and hybrid. The development approach applied differed by means of:
- project phases distinction and underlying activities organization,
- work planification and monitoring method.

The assignment, deliverables and grading scheme were independent of the treatment, hence shared by all teams.

#### 1) Iterative approach

In this approach there are no formal project phases as instead, certain steps are performed repeatedly. Such loops are called "iterations" and encompass different software development activities. In the context of this case study, an iteration lasts 2 weeks (the same as period between follow up classes), includes design, coding and testing activities and as its outcome, a piece of functionality is added to the target system. All of the above are captured in a "Definition of Done", which reflects the conditions to be met so that a given functionality can be considered complete. This practice implies that both the application and its documentation are delivered incrementally:

1. A wireframe corresponding to a given user story is created.
2. Source code is written and committed to a shared code repository.
3. HTML and PHP code quality is verified (see III.C).
4. Test case(s) linked to a given user story is defined.
5. Test case(s) is executed, and its outcome reflected in the Requirements Traceability Matrix.

Additionally, every iteration begins with a planning exercise and concludes with a demonstration of the developed part of the system to the course instructor. This is when formal feedback is provided; any suggestions or improvements can be then incorporated in an upcoming iteration. Partial design was not formally reviewed but feedback was provided upon request.

#### 2) Sequential approach

Sequential lifecycle models consist of a succession of phases, with no or little overlap between consecutive phases. In the context of this experiment, every project progressed through design, implementation and testing, each phase requiring specific artefacts to be produced. As part of the design, students were to conceive their solution and formalize it by means of documentation (as described in section II. A). The course instructor shared his feedback on the design before students could move to the next phase - implementation. It kicked off with a planning exercise for the whole phase ahead and onwards consisted mostly of coding activities with regular progress check points (formally monitored with KPIs). A final phase of testing consisted of executing prepared test cases and updating the Requirements Traceability Matrix to reflect their outcome. The last two weeks of the course were dedicated to bug fixing and polishing the application.

#### 3) Hybrid approach

Recent literature examined [10] how combining different methods is done with regards to the organization of the whole development process and identified three prevailing patterns. One of them - the Waterfall-Agile-Approach - uses the Waterfall model and its underlying phases as its baseline, however implementation is done in an iterative manner. It was chosen for this case study as it could be accommodated in the relatively short course timeframe (as opposed to the other two patterns). Similarly, to the sequential approach, the development process imposed progressing through dedicated design, implementation and test phases. As in the iterative approach, 2-week cycles were put in place during implementation and involved a planning exercise as well as a demo. The hybrid approach was additionally characterized by a practice concerning the team composition: "**align team structure with system architecture**" to help nurture technical skills and improve the internal quality of the system. It implied individual module ownership, meaning that a certain part of the solution was assigned to a single student. He was therefore responsible for all the aspects of that module: its design, implementation and quality assurance. The last part signifies that such student verified the quality of PHP and HTML code bases with corresponding

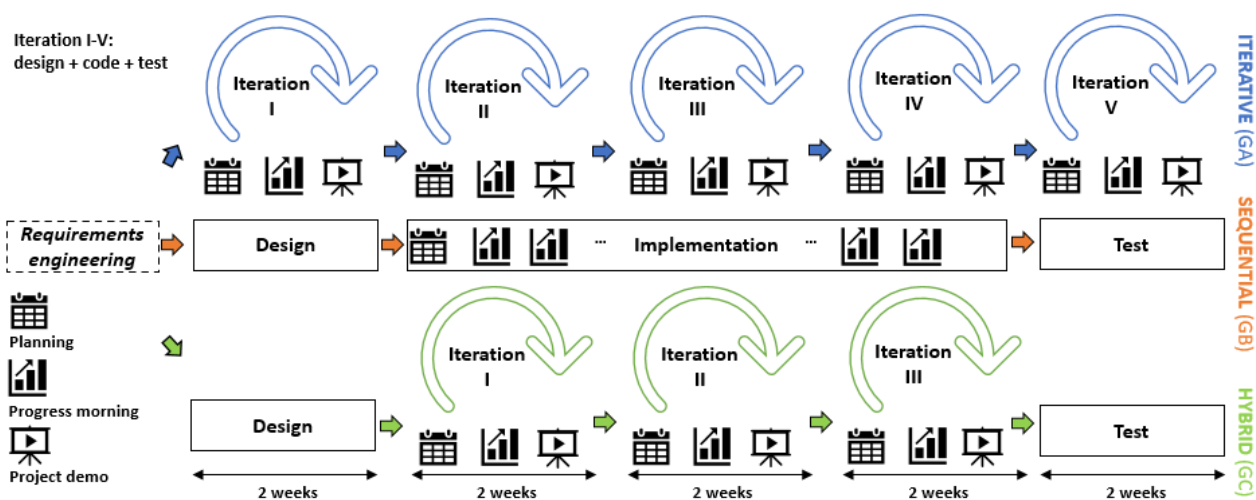

Fig. 2. A diagram presenting the structure of the three processes evaluated along with key student activities.

tools (see section III.C) and executed test cases covering underlying user stories.

As part of the project design, students had to specify the solution architecture with its underlying modules. That served as a basis for the distribution of responsibilities among the team members. Prior to starting the implementation, team members were requested to notify the course instructor about the functionalities under their responsibility via email.

### C. The study environment – student activities

This section describes activities that students had to perform as part of the course and details how they were contextualized for the approaches under study.

#### 1) Planning

Prior to starting any coding activity, all teams had to plan the work ahead (regardless of the process applied). It implied choosing the functionalities that will be implemented (their order suggested by the associated priority level, see II.A), identifying the underlying tasks and distributing them among the team members. Nevertheless, the scope of planning differed between the approaches. Students working in an iterative and hybrid manner were to plan work for the following two weeks whereas those working in a sequential fashion had to define tasks for the entire implementation phase (6 weeks) and set delivery targets for every follow up class. All students were required to formalize the scope of their upcoming work via an email sent to the instructor, listing all the functionalities planned in a given timeframe. Additionally, all tasks were logged in GitHub Issues and assigned to their owners.

As breaking down requirements into granular work items is not straightforward for novice developers [27], a sample user story and its translation into tasks was provided to the students. To further facilitate the planning exercise, the course instructor suggested to conceive the entire database scheme upfront and helped refine tasks granularity so that at least two layers are always distinguished (front end: HTML/CSS and backend: PHP).

#### 2) Work progress follow up

As form of the follow up and progress monitoring outside of classes, all teams were asked to organize a meeting every other week and as its outcome, the following items were addressed:
- update of the task statuses on GitHub issues (to do/ in progress/done),
- submission of a project summary report,
- calculation of work progress KPIs.

The first two activities looked the same for all groups. The meeting report format was imposed and took form of a table outlining the following items for each team member:
- advancement made since the last class,
- work to be done prior to the next class,
- problems encountered,
- risks identified.

On top of the summary, each group submitted a KPI showing the progress of work, calculated in a way that was characteristic to a given process. All teams received instructions and excel templates to be used for that purpose.

Students following the **iterative** approach used a simplified version of "Team velocity" to measure their productivity. Every week, they tracked the amount of completed tasks (those fulfilling the conditions of Definition of Done) versus the total number foreseen for a given iteration.

The **sequential** approach students employed one of the tools used to assess the project's performance in classical project management - Schedule Performance Index (SPI). It is a ratio of earned and planned value at a given point in time. In the study it was expressed by the number of tasks already tackled, divided by the total number of tasks identified during the implementation planning exercise. During every team meeting and follow up classes, students calculated the current SPI score and compared it with the target they have set initially (e.g 0.4 after 4 weeks of classes).

Students developing their solution with the **hybrid** method, made use of a metric originating from Lean manufacturing - Work In Progress - which corresponds to the amount of ongoing tasks at some point in time. A team was to maintain a limit on that number, which was established together with the instructor and was equal roughly to two tasks per team member. A Kanban board reflecting the possible statuses was available in GitHub and was used for tracking purposes. At the end of every team meeting, a summary of tasks distribution was handed in.

*3) Testing*

Sequential and hybrid approaches included a dedicated testing phase while the iterative way of working imposed inclusion of tests as part of every solution increment. Regardless of the process, all students had to perform the same tasks:

1. Write and document a test case(s) linked to a given user story.
2. Execute the corresponding test case(s).
3. Update the Requirements Traceability Matrix with the outcome of the test.

### D. *The study environment – roles & responsibilities within a team*

Every team had a designated Team Leader, who was there to guide the team towards the course's objectives, coordinate the work all along the project and be a point of contact for the instructor. Additionally, he/she was in charge of the following artefacts: team meeting summary and work progress KPI calculation.

Aside from Team Lead, no distinction of roles was made (e.g. tester or UX designer) and there was no notion of hierarchy - all the decisions were to be taken together (e.g. with regards to tasks distribution) and the team lead did not have authority over others.

### E. *Variables in the study*

**Independent variables**. The independent variable of interest in this study is the development approach applied by student teams. The effects of the following processes are evaluated: iterative, sequential, hybrid.

**Dependent variables.** The effectiveness of the software development method can be examined from different perspectives. In the study measures that encompass artefact, process and people facets of success were applied. A multi-dimensional approach for evaluation of the students' work, based on [11] inspired measurement of performance of a given processes in terms of the number of dependent variables.

- **Internal quality - HTML errors index:** given the Web nature of the assignment, the quality of client-side code was evaluated. A single score based on the number of errors generated during a W3C validation for a pre-selected portion of pages (unknown to the students), and divided by the total number of lines of code was used in this regard (1)

$$HE = E/n \quad (1)$$
*where: E - number of errors detected during W3C validation, n - number of lines in a HTML page*

- **Internal quality - HTML warnings index:** it is a measure similar to the first dependent variable, yielding the number of warnings per line of code (2).

$$HW = W/n \quad (2)$$
*where: W - number of warnings detected during W3C validation, n - number of lines in a HTML page*

- **Internal quality - maintainability ranking of PHP code:** this is a single maintainability measure that consolidates different technical aspects of the produced software, as defined by SIG, a software management consulting company, in collaboration with TV Informationstechnik laboratory. The underlying technical quality model is based on a number of metrics [12]:
    - Lines of Code (LOC),
    - duplicated LOC,
    - Cyclomatic Complexity,
    - parameter counts,
    - dependency counts.

- **External quality - Functional Correctness**: one of functional suitability characteristics defined in the ISO 25010 Product Quality Model [13] represents the degree to which a system provides the correct results. In the study, it is expressed as a ratio of functions containing bugs to the total number of functions tested.

$$FC = 1- \Sigma C/D \quad (3)$$
*where: C - severity of issues detected in a function, D - number of functions described in requirements specification*

- **Team productivity - Functional Completeness:** ISO defines it as a degree to which the set of functions covers all specified tasks and user objectives [13]. Simply put, it is a measure of the team's output, expressed by the amount of functionality delivered. It is calculated as a ratio of the number of missing functions detected during evaluation and the total number of functions described in the requirements specification.

$$TP = 1- A/B \quad (4)$$
*where: A - number of missing functions detected in evaluation, B - number of functions described in requirements specification*

- **Teamwork quality - team cohesion -** defined in literature as a "shared bond that drives team members to stay together and to want to work together" [14]. As stated in [15] team cohesion is highly correlated with project success and is critical for team effectiveness [16]" thus it was used as a proxy of teamwork quality in the study and was assessed using an adapted form of The Group Environment Questionnaire [17].

- **Learning outcomes - soft skills**: were evaluated using rubrics and encompass organizational and inter-personal skills. The following aspects of efficient collaboration within a team were measured:
    - *teamwork*: in terms of conformity with the team's pace of work and engagement towards the team's goals,
    - *planning*, in terms of conformity with the project plan decided by the group,
    - *tasks management*, in terms of respect of the tasks distribution,
    - *collective decision-making*, in terms of negotiation skills and facilitating an agreement at the team level,
    - *contribution to a positive working environment*, in terms of mutual support, diplomacy and goodwill.

- **Learnings outcomes - technical skills**: assessing the courses technical learning objectives: HTML, CSS, PHP and SQL programming as well as relational database management.

### F. Hypothesis formulation

Based on the findings reported by Wlodarski et al. [18] who compare plan driven and iterative approaches in the context of student computing projects two hypotheses were made about the evaluated processes:

1. Hybrid approach would yield higher team productivity in terms of Functional Completeness as compared to the sequential approach; similar results are expected for the iterative approach given that both deliver functionality incrementally.
2. Hybrid approach will produce software of higher external quality as compared to the iterative approach owing to its dedicated testing phase; similar results are expected for the sequential approach.

The team organization practice introduced as part of the hybrid way of working (and absent from the other approaches), could additionally bring the following benefits:

3. Hybrid approach would produce software of higher internal quality as compared to both iterative and sequential approaches, owing to the practice "Align team structure with system architecture" which implies responsibility of a given system module and execution of all process activities at a team member level.
4. Students working with the hybrid method will exhibit higher level of technical skills as compared to both iterative and sequential approaches, owing to the practice "Align team structure with system architecture".

### G. Data used in the study

The study is performed with data from 67 third year students of a Telecommunications program at a French engineering school. As part of it, they spend six semesters (from BAC+3 to BAC+5, or equivalently from BSc to MSc) mastering four thematic pillars: signal processing, digital communications, networks, and computer science. A majority of the students graduated from scientific "classes préparatoires" (which consist of intensive courses in mathematics, physics, chemistry as well as introductory classes to computer science) and therefore have a similar, beginner skill set and experience with coding.

The list of course participants was known in advance and they were allocated to three laboratory groups, based on alphabetical order of their surnames:

- GA (23 students): iterative approach
- GB (23 students): sequential approach
- GC (21 students): hybrid approach

Within one group, teams of 3 were formed; whenever not possible, teams of 2 were constructed instead and had a reduced backlog to implement in order to accommodate for the smaller team size.

## III. OPERATIONS

### A. Preparation

The design of the study was formalized prior to the start of the experiment in a protocol that was reviewed by all three authors. Similarly, all student materials, samples and questionnaires were prepared upfront and reviewed by the authors.

As one of the most important sources of variation in empirical software engineering studies is the skill level of subjects [28], all students were asked to fill in a demographics survey before the first classes. It probed their background in technologies/skills relevant to the course; all questions were mandatory. No student reported professional experience in any of the assessed skills or significant programming acumen, hence none of the team had a head start.

### B. Execution

The project part of the course lasted a total of ten weeks and was interlaced with a 2h follow up class per group every two weeks. The final delivery was planned two weeks after the last class.

Prior to the first supervised assignment work, a kick-off meeting with all participants was held. Students were informed of the experiment, given its protocol, and advised on the intent of investigating the effects of following a given development approach. Nevertheless, they understood that performance of different laboratory groups would not be compared to influence the grades. All teams within the study shared the same evaluation scheme which was fully transparent and communicated upfront. This included the developed system (assessed from the perspective of its internal and external quality as well as team productivity, see section II.C), the associated documentation as well as timely delivery of artefacts (meeting summary, KPIs etc.).

During the first class every group received a presentation of the method of work assigned to them to ensure all study participants share a baseline understanding of the treatment. Afterwards, teams were formed, where students within a laboratory group were randomly divided by three. The Team Leader role was filled on a voluntary basis. A manual comparison of teams' relevant technological acumen (based on a questionnaire distributed prior to the classes) was performed to ensure that there were no major gaps in skill sets among teams. Starting the third week of classes, due to COVID-19 pandemic and a national lockdown, all subsequent activities were carried out remotely using Discord as the main communication channel.

As part of the second follow-up class, an introductory presentation to the concept of maintainability was given as it was one of the quality aspects evaluated. The metric used in this regard maps the underlying quality model to 10 simple guidelines [10] to be respected when producing quality code. Students were encouraged to work continuously on source code quality, nonetheless the Maintainability ranking for grading purposes was only calculated when the projects were submitted.

Throughout the semester students were guided on how to correctly carry out associated activities by the course supervisors as well as with templates and samples were provided for every artefact.

## C. Data collection

To assess the dependent variables (see section II. C) data was collected throughout the semester and upon final hand-in.

*1) Recurrent data retrieval*

Temporal evolution of team cohesion was monitored and served as a barometer of teamwork quality. Systematic probing of the measure could potentially give insights into arising conflicts or periods when the teams struggled to meet course objectives. To evaluate team cohesion in the context of software engineering teams, use of The Group Environment Questionnaire was reported in the past [25, 26]. It is a set of questions reflecting on perception of collaboration and confidence on the project; its subset applied in the study is presented in Table 1.

Students were asked to periodically indicate their agreement with the above statements on a four-degree Likert scale. A decision to track team cohesion every two weeks was made as a tradeoff between timely and useful feedback and avoiding excessively burdening participants with its administration.

*2) End of term data retrieval*

Once the students submitted their projects, quality measures were derived, and the learning outcomes assessed. The client-side code was evaluated in terms of HTML errors and warnings indices (1,2) for a common set of five Web pages, that were not known to the students. An online W3C validator was used for that purpose.

The server-side code was appraised using a Maintainability ranking, calculated with a tool integrated at the level of a GitHub repository and made available by the metric creators.

Evaluating Functional Correctness (3) and Functional Completeness (4) of the solutions implied verifying that all the project requirements described in the Backlog were implemented and scrutinizing any inconsistencies with the desired behavior. Upon detection of any issues with a given functionality, a weight on a scale from 0 to 1 is assigned, based on instructor's best judgement. While this method of ranking bug severity is not extremely accurate, it was sufficiently reliable for the purpose of this study.

TABLE I. THE GROUP ENVIRONNENT QUESTIONNAIRE USED

| |
|---|
| Our team is united in trying to reach the goals of the course. |
| Our team members have a common vision for the project's future. |
| Our team would like to spend time together once the course is over. |
| Members of our team do stick together outside of the course-related activities. |
| I am happy with my team's level of desire to succeed. |
| My team gives me enough opportunities to demonstrate my abilities and skills. |
| I enjoy being a part of the social activities of this team. |
| For me, this team is one of the most important social groups to which I belong. |

The students were asked to evaluate their learning outcomes using four-degree rubrics that were distributed online once the course finished. A decision not to include a "middle" degree was taken as to avoid neutral response. Every participant could assess his technical as well as soft skills. Additionally, it was possible to assess their peers' ability to collaborate efficiently among 6 underlying skills - the evaluated soft skills were the same for self and peer-appraisal.

Finally, a dedicated questionnaire on every development approach applied was distributed. It consisted of roughly 10 closed questions to be answered with a Likert scale and 1 open question. It was administered to probe students' perception of the corresponding process organization and provide qualitative data to address hypotheses.

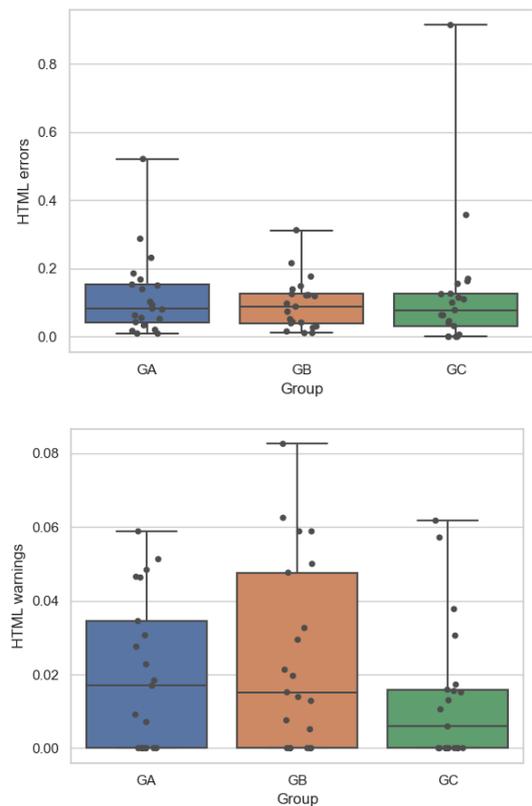

Fig. 2. Normalized number of HML errors (top) and warnings (bottom).

## IV. ANALYSIS AND INTERPRETATION

In this section we present the descriptive statistics and plots for each response variable and use them to analyze the results of the study. There are regrouped into three dimensions of success of student projects and presented accordingly: project quality, team productivity as well as human factors (teamwork quality & learning outcomes)

### A. Project quality

**Internal project quality** evaluation is carried out based on the source code that the team produced. From Fig. 2, it can be seen that the hybrid laboratory group had the best results in terms of a median, normalized number (per line of code) of errors and warnings. This observation could be explained by the practice linked to the hybrid method - "Align team structure with project architecture" - that asks every member to take on responsibility of a part of the project in its integrity, including the quality of the underlying code. Students' responses to the end of term questionnaire confirmed that they respected the practice (93.3% of positive answers). Nevertheless, only 40% agreed with the statement "I have verified the HTML code quality with W3C validator regularly" - the same score as observed for the sequential group (40,1%) but much higher than those of iterative students (18,8%). Overall results for the hybrid method could be considered as empirical evidence to link the team organization practice with a positive impact of internal project quality.

Findings relating to the second measure of internal project quality are not in line with the results concerning HTML code. Hybrid teams scored lowest with regards to PHP code quality in terms of the Maintainability rating (50.8%) while sequential and iterative groups demonstrated very similar results - 61,9%, 60,3% respectively. The lowest performance of the hybrid group can be linked to the fact that many teams following a hybrid approach did not work on that aspect of the project – only 20% of students confirmed regular use of the supporting tool - BetterCodeHub. That is a score lower than that of sequential (25%) and iterative teams (37,6%).

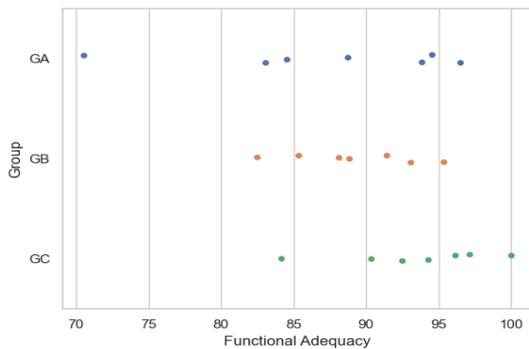

Fig. 3. Functional Adequacy - a measure of external quality.

**External project quality** was assessed using the Functional Adequacy (FA) metric, which is a measure of the severity of bugs (in terms of deviation from the project specification) detected in the final software. Fig. 3 presents scores for every team participating in the experiment, regrouped by the development approach used. What can be observed is that by excluding the outlier of GA (iterative teams), the FA results for both sequential and iterative approaches are almost identical. The hybrid teams exhibited the highest median (94.3%) and average (93.5%) values of FA, both scores being roughly 5% above the other two groups. This stems from the underpinning testing process of different approaches, which directly influences the Functional Adequacy.

Although all groups were expected to produce the same test deliverables, the means of execution differed between the approaches. Iterative teams were to test the delivered functionalities as part of Definition of Done and reflect the outcome in the Requirements Traceability Matrix (RTM) gradually; for sequential and hybrid groups it was a one-off effort as part of a dedicated testing phase. The above could imply that for iterative teams errors could have been introduced along with new functionalities and possibly teams did not re-test the entire solution before the hand-in. Secondly, the degree of respect of the conditions of Definition of Done varied among teams and reflected their strive for high quality - the RTM was assessed only at the end of the semester hence potentially not all teams consistently tested the functionalities delivered at the end of an iteration. These hypotheses were addressed in a questionnaire, where a large majority of students stated having rigorously tested requirements before the increment delivery (87.5%) and another round of tests before the final hand-in (93.7%). The highest values of hybrid teams can be explained by the fact that it combined both an incremental delivery and a dedicated verification phase. The first practice ensured more value was provided for the last follow up classes (five out of seven groups reported delivering 100% of the functionality in RTM while only three did so for the sequential group) thus more functionality was tested systematically and in turn the final deliverable demonstrated higher external quality.

### B. Team productivity

In a professional context, team productivity relates to resources utilization and efficiency [19] and as a consequence is tracked by metrics measuring teams' efforts in terms of time and development estimations. However, the number of logged hours did not prove to be a representative measure of team productivity in academic projects [20, 21], thus in our experiment the output of students' work is used as a proxy in this area instead. Functional Completeness (FC) of all teams, which represents the ratio of functional requirements delivered and the total number described in the assignment specification, is depicted on Fig. 4.

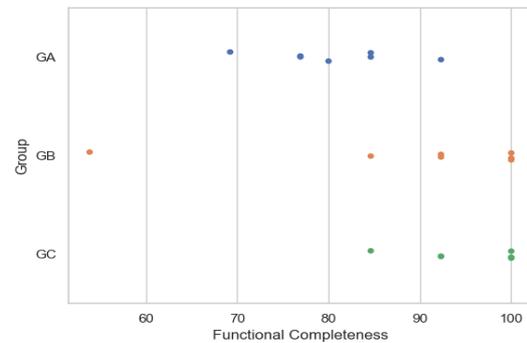

Fig. 4. Functional Completeness – a measure of team productivity.

It can be clearly seen that the median value for the iterative approach (80%) is much lower than that of sequential (92%) and hybrid way of working (100%). What probably greatly contributed to the final score of FC is an intermediary milestone present for the latter two approaches. Two weeks prior to the final deadline, the testing phase started and during the last follow up classes, students were expected to execute defined test cases on a nearly complete solution. This naturally shifted the peak of their implementation efforts earlier, so that they could test a semi-finalized project as requested by course instructors. Iterative teams, despite a regular, partial hand-in could have had an impression of more time due to a distant deadline for final delivery (which was the same among all laboratory groups). As a consequence, they did not advance swiftly enough throughout the semester and delivered the smallest number of functionalities (apart from the highest performer in the group that finished off the application two weeks before the end of the classes, however disregarded some of the functions thus yielding a score of 93%).

While results of Functional Completeness for sequential and hybrid approaches seem very similar, the latter scored slightly higher in terms of the median value (even when excluding the outlier for the sequential group) as all teams except one delivered 93% of requirements or more. Similar to the outcomes of the project in terms of Functional Adequacy, this phenomenon can be explained by the iterative nature of coding activities. They resulted in a higher degree of project completion prior to the testing phase, thus leaving less work for the last two weeks and increasing chances of handing in a finalized assignment.

## C. Human factors

Teamwork quality can have a great impact on the end-result of a group undertaking, therefore effective team functioning is cited as one of the success factors in project management [22]. In order to consider a university course successful, students' learning outcomes should be examined. Consequently, both aspects are taken into consideration in the experiment to evaluate the impact of development approaches under study on the success of the projects from a human perspective.

### 1) Teamwork quality

Longitudinal data on team cohesion, which is taken as a proxy of teamwork quality, are presented on Fig. 5. The evaluation of its perception probed on a four-Likert scale, was regrouped into positive and negative impressions, and represented on the graph as a function of course progression.

What can be seen is that all teams started off with a very similar level of team cohesion (in the range of 82.5-85%), after two weeks of group work. The metric then degrades over time for the iterative and sequential approaches. The evolution of team cohesion level for these groups resembled a downward spiral which probably reflected more and more deviation from the target progress which naturally lowers students' satisfaction with teamwork. It should be noted that two teams working in an iterative manner reported issues (health problems and uneven contribution to the project respectively) to the course instructor. These could have had a negative impact on their perception of team cohesion lowering the scores for the whole study group as a result.

The hybrid approach did best over time, despite a sharp dip (and lowest score from all groups) at the end of the first iteration. From then on, the team cohesion level bounced back, surpassing the initial high while the other two groups (iterative and sequential) continued the decline. Explanation is that week five marked the first demo for the hybrid teams hence extra efforts were needed to integrate different components and finish off something demonstrable. Sequential teams also exhibited the highest decrease in team cohesion that week (as compared to other weeks), suggesting that the first two weeks of implementation efforts were rocky for all teams who had to switch from design to coding activities (hybrid and sequential). The end of term questionnaire supports this hypothesis as 33,4% and 22,3% of students in hybrid and sequential groups respectively reported difficulties in switching from design to implementation as part of a formal phase transition.

### 2) Learning outcomes

A total of 10 course learning objectives were evaluated using a rubrics system: five technical and five soft skills.

TABLE II. PERCEPTION OF TECHNICAL SKILLS ACQUISITION. 4-LIKERT RESPONSES WERE REGROUPED IN TWO CATEGORIES: POSITIVE AND NEGATIVE IMPRESSIONS

|  | **Iterative (GA)** | **Sequential (GB)** | **Hybrid (GC)** |
|---|---|---|---|
|  | **HTML** | | |
| POSITIVE | 100.00% | 100.00% | 94.74% |
| NEGATIVE | 0.00% | 0.00% | 5.26% |
|  | **CSS** | | |
| POSITIVE | 61.54% | 82.35% | 73.68% |
| NEGATIVE | 38.46% | 17.65% | 26.32% |
|  | **BDD** | | |
| POSITIVE | 84.62% | 94.12% | 89.47% |
| NEGATIVE | 15.38% | 5.88% | 10.53% |
|  | **SQL** | | |
| POSITIVE | 100.00% | 94.12% | 94.74% |
| NEGATIVE | 0.00% | 5.88% | 5.26% |
|  | **PHP** | | |
| POSITIVE | 84.62% | 94.12% | 94.74% |
| NEGATIVE | 15.38% | 5.88% | 5.26% |

| **OVERALL** | **Iterative (GA)** | **Sequential (GB)** | **Hybrid (GC)** |
|---|---|---|---|
| POSITIVE | 86.15% | 92.94% | 89.47% |
| NEGATIVE | 13.85% | 7.06% | 10.53% |

Results collected for different facets of effective collaboration (which are described in more detail in section II.D) are presented on Figure 6. What can be observed is that the hybrid teams scored highest in all of the evaluated aspects of conditions of collaboration, consistently securing the highest share of soft skills rated as "excellent". For most of the categories, the sequential group came in second and the iterative third. Similarly to the results of team cohesion, the low scores of iterative teams could be an aftermath of the health and work distribution issues reported by two teams during one of the classes.

TABLE III. STUDENTS' VIEW OF THE IMPACT OF A GIVEN PROCESS ON SOFT AND TECHNICAL SKILLS DEVELOPMENT

|  | *Strongly disagree* | *Disagree* | *Agree* | *Strongly agree* |
|---|---|---|---|---|
| **The development approach applied during the project, helped me to collaborate efficiently with the team** | | | | |
| GA | 0% | 31.3% | 37.5% | 31.3% |
| GB | 11.8% | 29.4% | 41.2% | 17.6% |
| GC | 0% | 6.3% | 31.3% | 62.5% |
| **The development approach applied during the project, helped me developed technical skills in Web programming (PHP, HTML)** | | | | |
| GA | 0% | 6.3% | 37.5% | 56.3% |
| GB | 0% | 17.6% | 41.2% | 41.2% |
| GC | 0% | 0% | 31.3% | 68.8% |

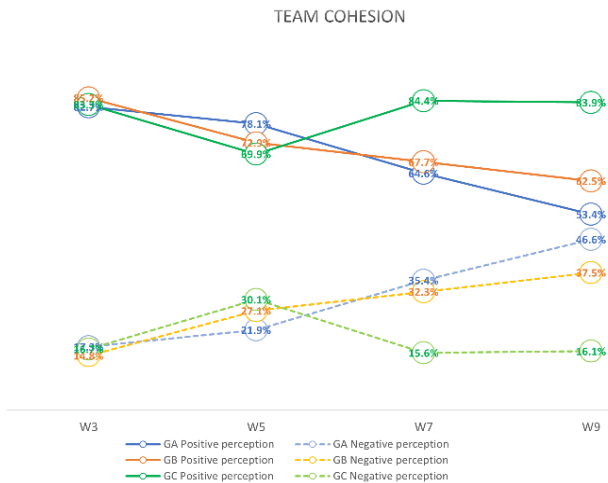

Fig. 5. Temporal evolution of team cohesion a proxy for teamwork quality.

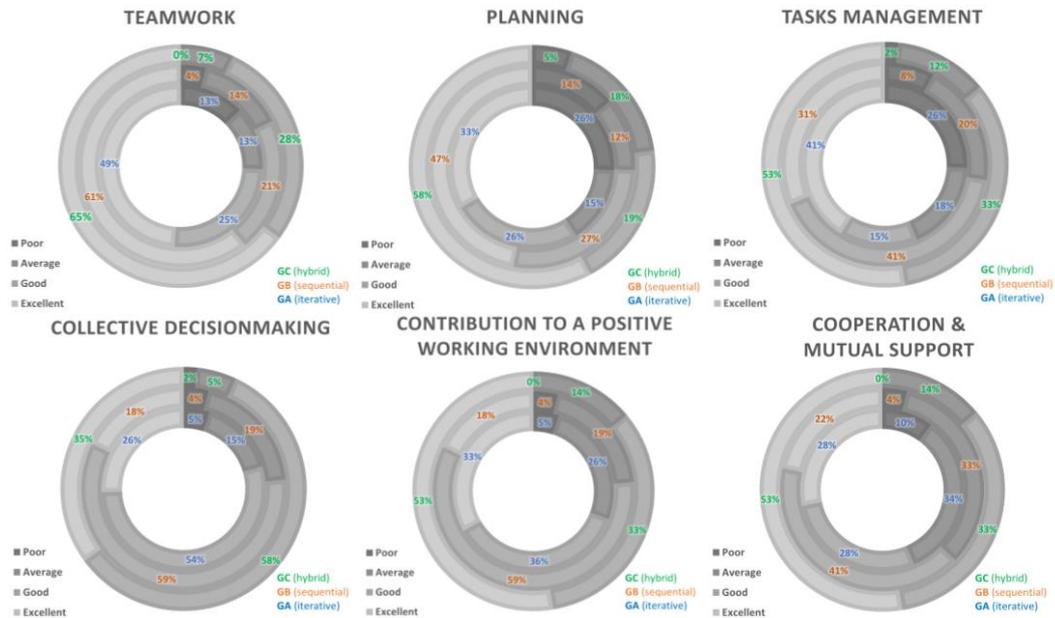

Fig. 6. Acquisition of soft skills reported by students using 4-degree rubrics.

Acquisition of technical skills is based on a self-evaluation of every student, thus much higher rates of positive answers are reported as compared to soft skills (which combines self and peer assessment) - refer to Table II. Each development approach recorded the highest value at least once but there are no consistent patterns across the competencies probed. Furthermore, the differences among the average values of all skills per group are negligible - a spread of less than 7% can be observed between highest and lowest performer. Given the variation in the findings, a link between a development approach and acquisition of technical skills cannot be established.

When it comes to students' judgement of a given software development process and its impact on the skills acquisition, the hybrid approach scored highest for both soft and technical skills. A summary of the results is presented in Table III.

It is important to note that no statistical analysis of the results was performed due to a relatively small number of data points.

## V. THREATS TO VALIDITY

As empirical research, this study is subject to different types of threats - they are described according to the classification suggested by Wohlin et al [23].

A comprehensive, metrics-based approach to the evaluation of the experiment, which in principle limits bias and uncertainty in the assessment process, ensures a relatively high **internal validity**. Furthermore, random assignment of students among groups is a suitable way to distribute study sample in a controlled experiment. Nevertheless, there is a risk that project success is influenced by uncontrolled factors other than the development approach used, such as varying student affinity or motivation. While these aspects were not accounted for, they are inherent to a university setting hence should not pose a problem as an internal threat.

Concerning the **external threats**, it is highly probable that comparable results should be obtained when running the same course in subsequent years. Likely a similar project assignment and class set up at a different university would yield comparable results, granted a balanced distribution of skillsets across the teams. However, we have no arguments to assess the generalizability of setting - impact of a given development approach can vary for more experimented students, larger team size, longer courses and other Information Technology domains that Web programming.

**Construct validity** includes a major threat concerning the teamwork quality assessment. Team cohesion is just one of many measures of conditions of collaboration in teams, thus provides a partial picture of the phenomenon. However, in order not to excessively burden participants with surveys, no other teamwork quality facets were assessed.

Regarding the **conclusion validity**, some of the tools used in the experiment suffer from a certain degree of subjectivity. Functional correctness metric mirrors the evaluator's judgement of severity of the detected anomalies, whereas team cohesion questionnaires and skills assessment rubrics fully rely on students' shifting perceptions of collaboration on a given day and can be impacted by team conflicts, which are not process-bound. To systematize the external quality verification, all anomalies detected were noted down so that whenever a similar problem was detected in any other solution, the same score would be given. To avoid bias in feedback collected via questionnaires/rubrics, it was recalled that they were not considered for the grading scheme. The intention was to avoid skewed responses giving a positive impression of team cohesion or their/peers' skills.

The reliability of treatment implementation was addressed by an appraisal of team conformance to a given development approach. Indeed, timely delivery of all expected artefacts (project code; documentation, meeting summary, work progress KPI, demo etc.) was tracked and a mark was given to every team, which contributed to the final grade for the project. On top of that, special efforts were made to avoid introducing bias to any of the laboratory groups, as more than one instructor intervened during the experiment (iterative and hybrid teams shared the same supervisor, whereas the sequential ones were accompanied by another professor). Instructor's role in every class activity was detailed prior to the experiment and documented in its protocol. For example, as part of the hybrid method, the course instructor provided his feedback only on the design artefact and iteration demos were meant as checkpoint on the progress rather than sharing enhancement suggestions (as opposed to the iterative approach, where there was no formal design review and feedback on the effects was provided during demos). Likewise, all the teams received the same guidance and level of instructions as well as samples with regards to planning, tasks identification and testing.

## VI. CONCLUSIONS

With this work, we aimed to investigate how the choice of an iterative and sequential development method influences success of student team computing projects as contrasted to a hybrid way of working that combines both approaches. Three axes of evaluation that encompassed 16 metrics bring forward certain patterns and provided quantitative data to answer the research question raised.

Indeed, the first hypothesis turned out to be true as the use of the hybrid approach contributed to a considerable improvement of team productivity in terms of the number of functionalities handed in - all teams except one delivered 93% of requirements or more. According to our observations, this success was mostly due to the introduction of a dedicated testing phase. It effectively advanced the peak of coding activity as compared to the iterative approach, and in turn more functionality was completed two weeks before the final deadline. Both sequential and hybrid approaches scored significantly higher in terms of Functional Completeness as compared to the iterative one. The hybrid way of working, which incorporated incremental delivery, helped to bring about a further gain in team productivity as compared to the sequential development. This was reflected by higher scores of system completion during the last follow up class - they were reported by students in the tests results (reflected in the Requirements Traceability Matrix).

Moreover, the external quality of the projects measured in terms of Functional Adequacy (FA) seems to partially confirm the second hypothesis. The hybrid teams scored highest in that regard, approximately 5% more than the other two approaches. It seems that having a dedicated testing phase is not the only success factor here given that the FA values were very similar for iterative and sequential approaches: median of 88.8 vs 88.9 and average of 87.4 vs 89.0 respectively. It is a mix of iterative development and a formalized testing phase that yield the best results in our study.

Based on the data collected, it is difficult to address the two hypotheses concerning the practice "Align team structure with system architecture" (characteristic of the hybrid approach) which was meant to support the technical skills acquisition and positively impact the code quality. While the teams working with that method scored highest in terms of HTML code quality, the differences between the groups were relatively small. Furthermore, the hybrid teams scored lowest when it comes to PHP code quality which was due to the failure to use the supporting tool (56.3% of students in the group reported never using it).

In the study we also considered the impacts of a development approach from a human perspective - the teamwork quality as well as acquisition of soft and technical skills. The highest levels of team cohesion for hybrid teams were observed; additionally, it was the only group not to experience a decline as compared to the beginning of the project. Positive perception of teamwork quality for hybrid teams was further confirmed by the end of term questionnaire where students were to evaluate the evolution of team collaboration over time. There were 93.3% of positive impressions of it among hybrid students, 62.5% among iterative team members and 58,8% among the sequential ones. The low team cohesion levels exhibited by the iterative group seem to counter the human-centric angle of the agile way of working reported in the past [24, 25, 11]. This could be partially explained by the fact that two iterative teams reported health and work distribution issues to the course instructor. Nevertheless, teams in sequential and/or hybrid group could have experienced similar obstacles without voicing them.

Finally, working with the hybrid approach brought about a much more positive perception of soft skills acquisition among the team members. These students reported consistently higher values than those working with other approaches in the rubrics that assessed six facets of efficient collaboration. While it is difficult to establish a direct link with a specific aspect of the hybrid way of working, students that followed it exhibited the highest values of most of the "human-centric" variables tracked (that excludes technical skills acquisition, where inconsistent results among groups were reported)

In the study we presented the impact of different types of methodologies on the product, project and people involved. While the methodologies imitate industrial practices, our observations cannot be extended to a professional setting without further research. Instead, the patterns observed serve as a blueprint for other educators who wish to tailor or design a software development process for academic needs and guide students' computing projects to help ensure a positive outcome of the project. We therefore recommend putting in place certain practices, such as incremental project delivery coupled with formalized quality assurance – as applied by the hybrid teams – to safeguard high team productivity without quality shortcomings. Despite anecdotal empirical evidence, we encourage introducing the practice "align team structure with system architecture". We believe that each member's full accountability of a given system module and thus a clear distribution of work could nurture healthy relations among peers and enable efficient collaboration. Furthermore, regular, and formalized communication put as part of the study – team status meetings and progress tracking (see II.C 2) – can address challenges in this area, which stem from busy schedules, planning issues and lack of experience and training [24]. Guidelines described in the paper could prove particularly useful in the context of remote learning.

Ultimately, exploiting a process adapted to the academic setting to guide students in building an entire operational software can help address the discrepancies between the skill set of graduates and employment needs. A recent study [26], explicitly lists software design and testing among the most important deficits in Computer Science education. Thus, providing students with well-defined hands-on experiences building a software system through design, implementation, testing and management activities could be a steppingstone to bridge that gap.

ACKNOWLEDGMENT

We would like to thank Bordeaux INP for their support through the grant "AAP Initiatives Pédagogiques 2020".